\documentclass[12pt]{article}
\usepackage[english]{babel}
\usepackage{epsfig}

\begin{document}
\title{\Large \bf Fermion pair production at LEP2 and interpretations}
\author{G.~Abbiendi \bigskip\\
{\it INFN Bologna, Italy} \\
{\small {\it E-mail:} \sf Giovanni.Abbiendi@bo.infn.it}}
\date{}
\maketitle
\vspace*{-0.8cm}
\begin{center}
{\footnotesize To appear in the {\it Proceedings of the New Trends in 
High-Energy Physics}\\ Yalta, Ukraine, September 22 - 29, 2001.}
\end{center}
\vspace*{0.1cm}
\begin{abstract}\vspace*{0.1cm}
Preliminary results on $e^+ e^- \to f \bar{f}$, $f=e, \mu, \tau,
q$, including all LEP2 data 
are discussed. 
Good agreement is found with the Standard Model up to the highest energies.
Limits on possible new physics are extracted.
\end{abstract}
\vspace*{0.4cm}

Fermion pair production processes 
have been measured by the four LEP experiments up to 
$\sqrt {s} \simeq 207$~GeV \cite{lepew}.
Above the $Z$ peak $\gamma$ radiation is very important, leading in particular 
to a high rate for the $Z$ radiative return. Events can be classified according
to the effective center of mass energy $\sqrt {s'}$, 
which is measured in different ways. A typical inclusive selection requires 
$\sqrt{s'/s} \geq 0.1$, while events with only a low amount of radiation ({\em
exclusive}) are defined by $\sqrt{s'/s} \geq 0.85$. Exclusive events are
obviously more relevant to look for new physics.
The signal definition is complicated by initial-final state interference.
Two theoretical definitions have been considered for
$\sqrt {s'}$ in the combinations of LEP data:
\begin{enumerate}
\item 
$s$-channel propagator mass, with interference between initial and final
state radiation subtracted (used by L3 and OPAL);
\item 
bare invariant mass of the dilepton pairs, or $s$-channel propagator mass
for hadronic final states, with interference included (close to ALEPH and DELPHI
definitions).
\end{enumerate}
For $e^+ e^-$ pairs $\sqrt{s'}$ is not natural, as $t$-channel exchange diagram
dominates: in this case {\em non-radiative} events are selected by a cut on
the acollinearity angle of the final state electrons, typically $\theta_{acol}
\leq 10^o$.
Another delicate point is the contribution from 4-fermion processes which enter
the pair selection,
which has to be defined by a cut on the invariant mass of the extra pairs.

Theoretical uncertainties have been assessed during the LEP2 MC Workshop 
\cite{lep2mc}
and are presently well below the experimental errors for $q \bar q$, 
$\mu^+ \mu^-$ and $\tau^+ \tau^-$ pairs. They amount respectively to $0.26~\%$
for $q \bar q$ and $0.4 \%$ for $\mu^+ \mu^-$ or $\tau^+ \tau^-$ cross sections.
On the opposite side, the theoretical uncertainties on Bhabha cross sections are
still large, $2 \%$ in the barrel region and $0.5 \%$ in the endcap regions: 
a sizeable reduction (factor $4$-$10$) 
is desired to exploit the experimental precision.

\begin{figure}[tb]
 \begin{center}
   \mbox{
     \includegraphics[width=0.4\linewidth]{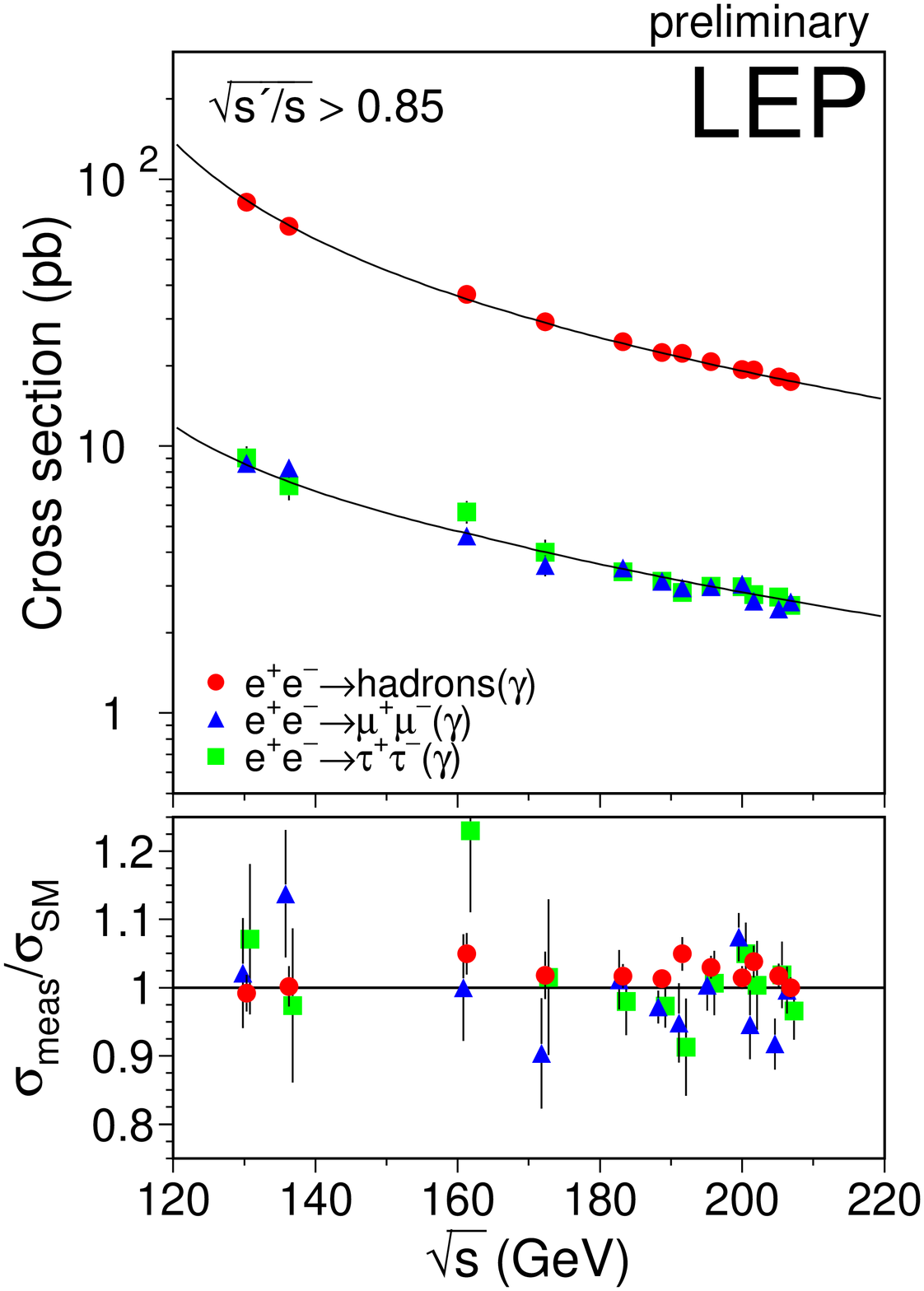}
     \includegraphics[width=0.4\linewidth]{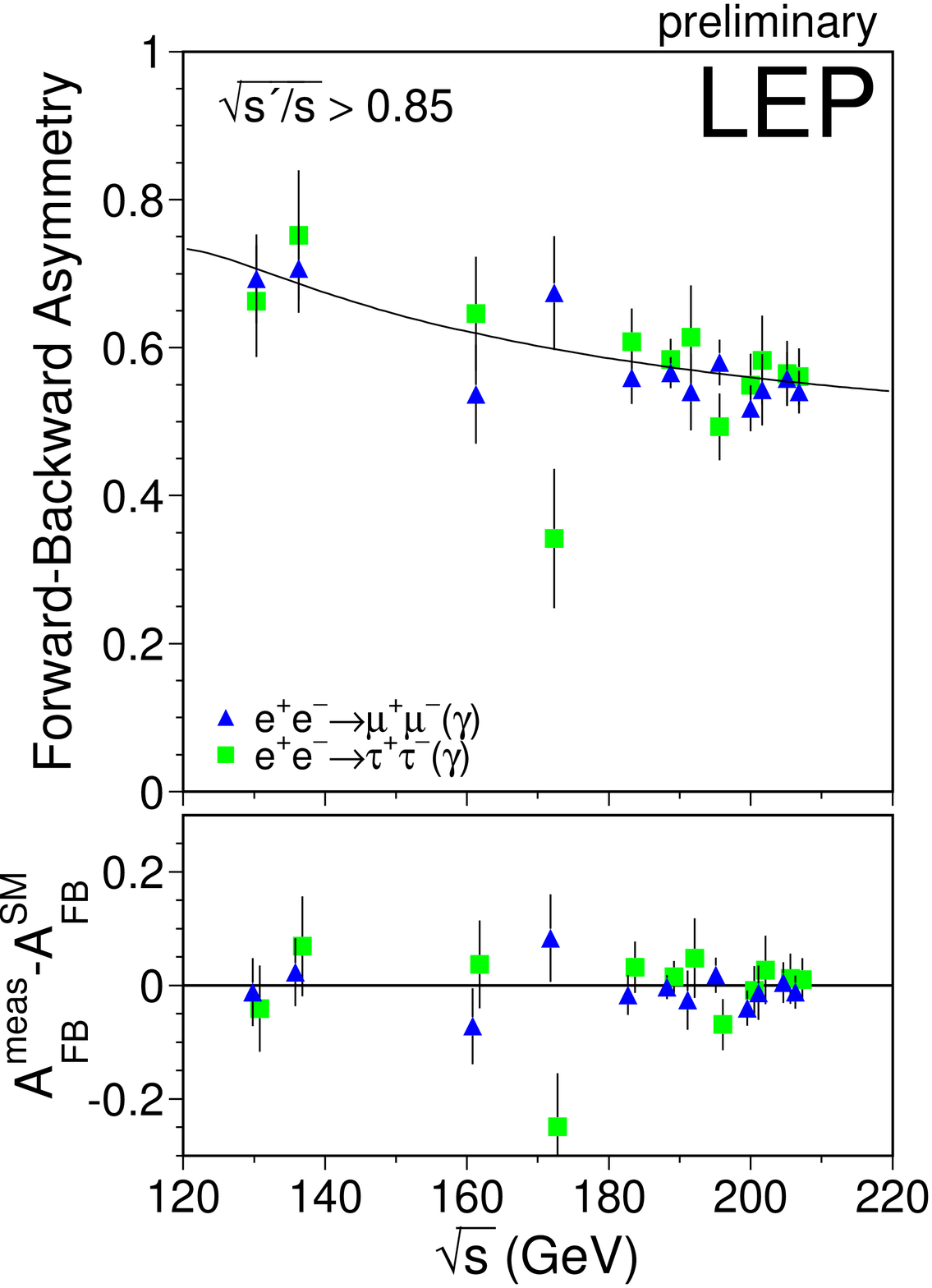}
     }
 \end{center}
\caption{\small 
Cross sections for $q \bar q$, $\mu^+ \mu^-$ and $\tau^+ \tau^-$ final
states and forward-backward asymmetries for $\mu^+ \mu^-$ and $\tau^+ \tau^-$ as
a function of energy. The curves are SM predictions.}
\label{sigmas}
\end{figure}

Preliminary combinations of LEP data exist 
for the exclusive $q \bar q$ cross sections and for 
$\mu^+ \mu^-$ and $\tau^+ \tau^-$ cross sections and forward-backward
asymmetries over the whole energy range ($130$-$207$~GeV) \cite{lepew}, 
as shown in figure \ref{sigmas}.
Standard Model (SM) expectations are obtained with ZFITTER \cite{zfitter}. 
Correlations within/between experiments have been taken into account
in the combinations.
The combined errors are dominated by
statistics and uncorrelated systematics.
Moreover differential cross sections $d \sigma / dcos\theta$ 
have been combined for $\mu$ and $\tau$ pairs for 
$183 \leq \sqrt{s} \leq 207$~GeV.
Available heavy flavour measurements of $R_b$, $A_{FB}^b$, $R_c$, $A_{FB}^c$
have been combined at all LEP2 energies \cite{lepew}.
Bhabha measurements have not been combined yet, 
though each experiment has a complete set of measurements, see for example
\cite{opalpn} or the references in \cite{lepew}.
All the LEP averages are in good agreement with the SM predictions, as
each experiment's results. Therefore such data have been used to set indirect
limits on a number of new physics scenarios.

An alternative test of the Standard Model is possible in the
$S$-matrix approach \cite{smatrix}. 
In this framework the only assumptions are the existence of
a heavy neutral boson ($Z$) in addition to the $\gamma$ and validity of QED 
for photon exchange and radiation.
In particular $\gamma$/$Z$ interference is left free, 
while it is usually constrained by the SM itself in fits
of the $Z^0$ lineshape.
LEP1 data have low sensitivity to
$\gamma$ exchange and $\gamma$-$Z$ interference. An $S$-matrix fit restricted 
to LEP1 data shows a strong correlation between the fitted mass $m_Z$ and the
parameter $j_{had}^{tot}$
related to $\gamma$-$Z$ interference in the hadronic cross section. 
In a L3 analysis \cite{l3_1}
such correlation brings about an additional $\pm 9.8$~MeV uncertainty to $m_Z$.
LEP2 data strongly constrain $\gamma$-$Z$ interference terms. 
L3 \cite{l3_2}
fitted jointly all LEP1+LEP2 cross section and asymmetry measurements,
either assuming lepton universality or not.
The result is $m_Z = 91188.4 \pm 3.6$~MeV 
($m_Z = 91188.8 \pm 3.6$~MeV without lepton
universality), in agreement with the SM lineshape fit. 
Here the correlation is reduced and contributes an error of 
$\pm 1.8$~MeV, already included in the quoted result.
The fitted value of $j_{had}^{tot}$ is
$0.30 \pm 0.10$, in agreement with the SM prediction of $0.21$.
Similar results have been obtained by OPAL \cite{opalsmat}.

\begin{figure}[tb]
\begin{center}
\hspace*{-1cm} 
   \mbox{
\epsfig{file=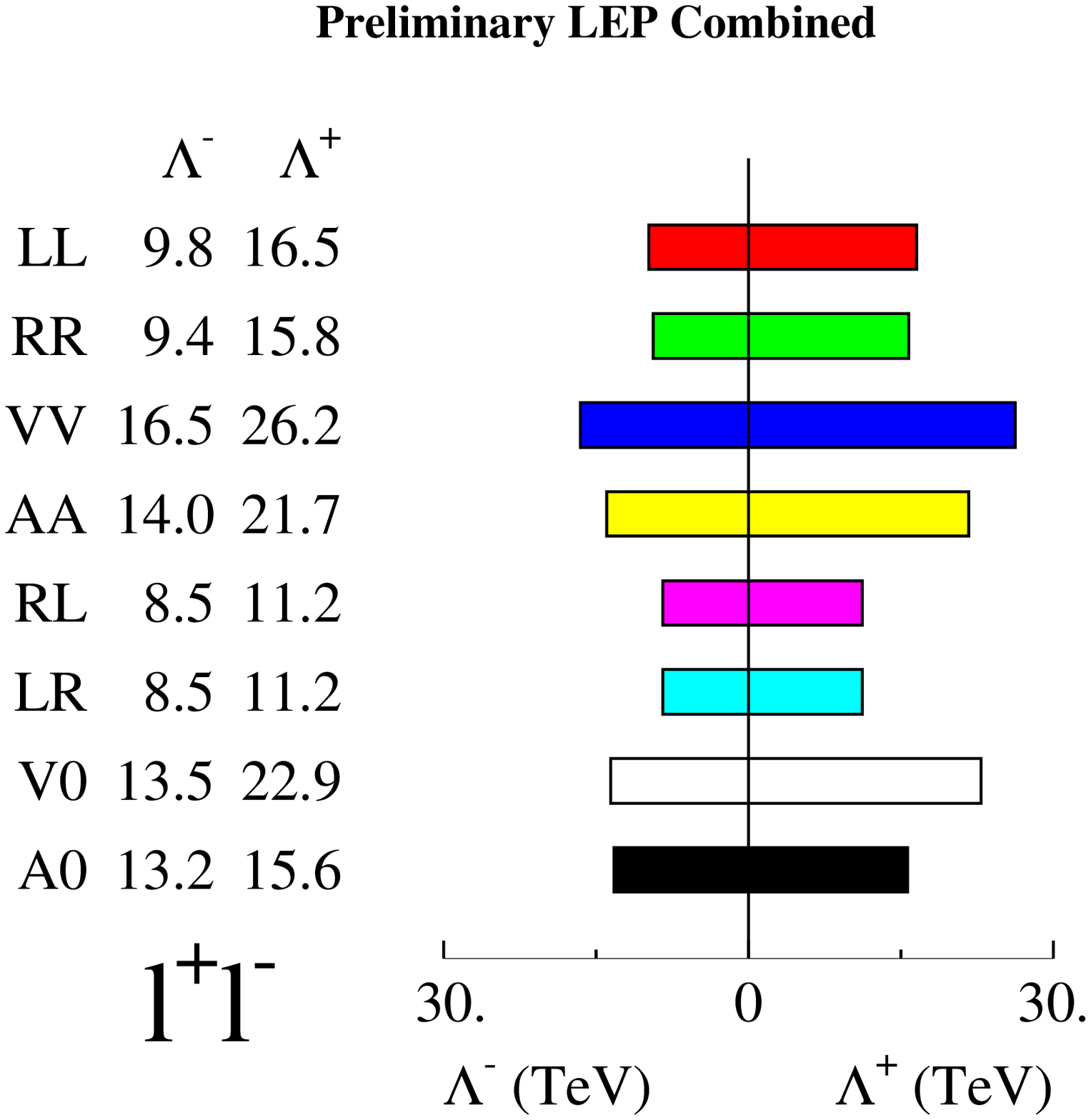,height=7cm}
\hspace*{-1cm}
\epsfig{file=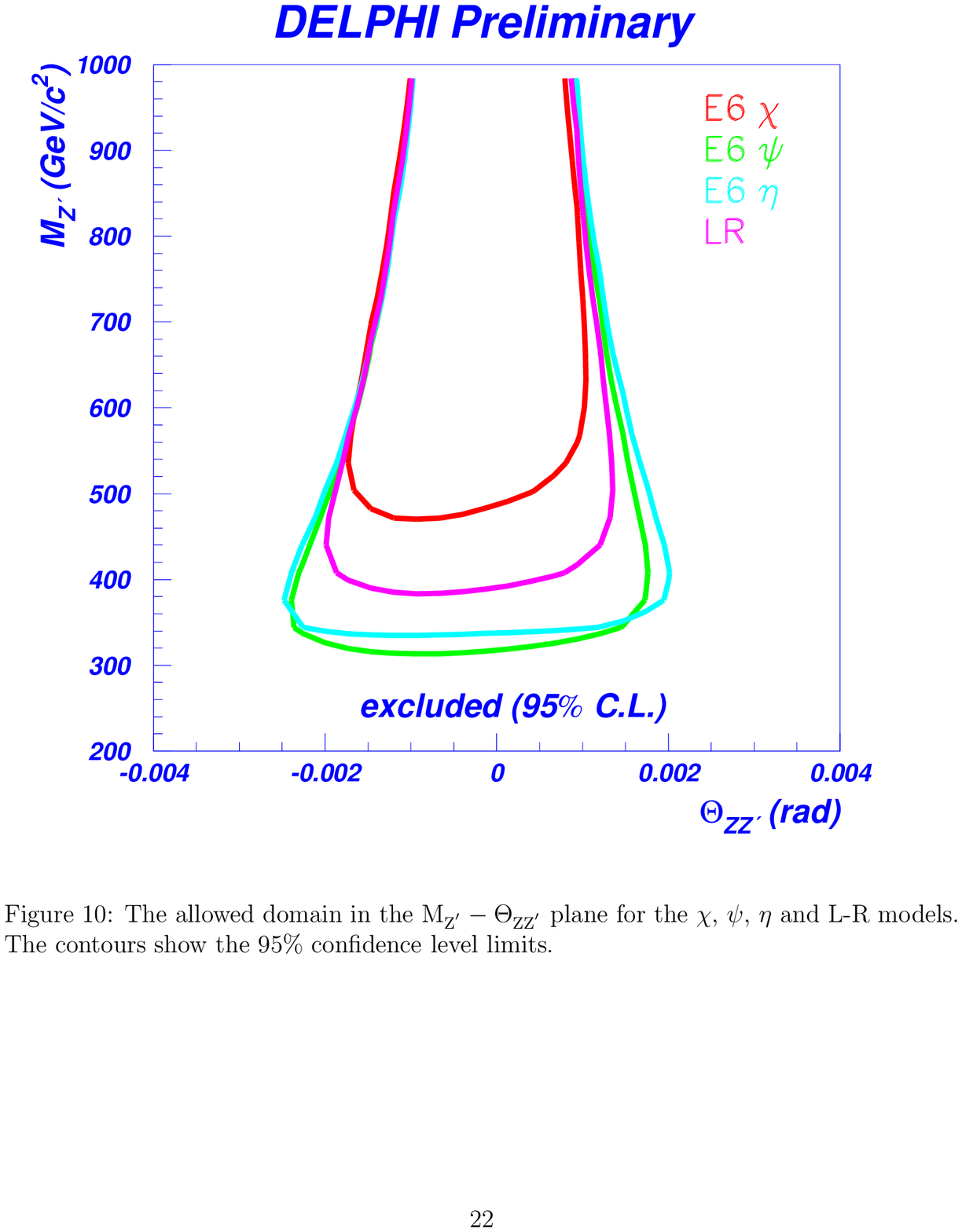,height=6.5cm
,bbllx=71pt,bblly=235pt,bburx=525pt,bbury=632pt,clip=
 }
        }
\end{center}
\caption{\small 
{\it (Left)} 
Limits on the scale $\Lambda$ of $eell$ contact interactions assuming
$\mu/\tau$ lepton universality;
{\it (Right)}
Exclusion contours in the $M_{Z'} - \theta_{ZZ'}$ for some GUT models.
}
\label{figlimits}
\end{figure}

A convenient way to describe any deviation from the SM 
in $e^+ e^- \to f \bar f$ is the framework of four-fermion contact
interactions \cite{4f}, which is appropriate if the scale of new physics 
$\Lambda$ is much greater than $\sqrt{s}$.
LEP averages of $\mu^+\mu^-$ and $\tau^+\tau^-$ cross sections and asymmetries
have been used for such indirect search. 
They give at present the best lower limits on the scale $\Lambda$ 
for $eell$ contact interactions,
in the range of $8.5$ to $26.2$~TeV depending on the specific model
($95 \%$ C.L. limits assuming conventionally a strong coupling $g^2 = 4\pi$)~\cite{lepew}.
In detail the limits for each model and both signs of interference
between the hypothetic new interaction and the SM are shown in 
Fig.~\ref{figlimits}~({\it left plot}). 
Furthermore LEP combinations of heavy flavour measurements 
have been used to set lower limits on $eebb$ and $eecc$ contact interactions. 
Depending on the model they are in the range of $2.2$-$14.6$~TeV
for $eebb$ and $1.4$-$7.4$~TeV for $eecc$~\cite{lepew}. 

Limits on the masses of new heavy particles have been extracted also 
within specific extensions of the SM. This is actually
appropriate when the mass of the new particle is of the same order of
magnitude as the center of mass energy. 
New particles coupling to leptons and quarks could be 
leptoquarks \cite{buch}
or squarks in supersymmetric theories with R-parity violation.
Beyond the kinematic limit for direct
production, they could be observed through a change of the total
cross section and asymmetry 
in the process $e^+ e^- \to q \bar q$ via a $t$-channel exchange diagram
\cite{kali1}.
The best LEP limits come presently from ALEPH \cite{aleph2001}. 
They have been extracted separately for leptoquarks/squarks of each
of the three families, profiting of $b$-tagging and jet-charge techniques.
It is assumed only one new particle contributing at a
time, with coupling only to left or right-handed leptons.
In particular
the mass limit for $S_0(L)$ coupling to first or
second generation quarks (equivalent to $\tilde{d}$ or $\tilde{s}$) 
is about $600$~GeV, 
for ${\tilde {S}}_{1/2} (L)$ coupling to third generation quarks
(equivalent to $\tilde{\bar t}$) is about $140$~GeV ($95 \%$ C.L. limits
assuming electromagnetic strenght for the coupling $g^2 = 4 \pi \alpha$).
They are complementary to limits obtained from HERA, Tevatron, and
low energy data (atomic parity violation, rare decays).

Supersymmetric theories with R-parity violation have terms in the Lagrangian of
the form $\lambda_{ijk} L_i L_j \bar{E_k}$, being L a lepton doublet superfield
and E a lepton singlet superfield. The parameter $\lambda$ is a Yukawa coupling
and $i$, $j$, $k = 1$, $2$, $3$ are generation indices. For dilepton final
states, both $s$ and $t$-channel exchange of R-parity violating sneutrino can
occur \cite{kali2}.
The strongest limits are obtained when $s$-channel resonant production of 
${\tilde{\nu}}_\mu$ or ${\tilde{\nu}}_\tau$ is possible. This could be detected,
depending on the non vanishing couplings, in the 
$e^+ e^-$, $\mu^+\mu^-$ or $\tau^+\tau^-$ decay channels.
Dilepton differential cross sections have been used by ALEPH 
to set upper limits on the couplings as a function of the sneutrino mass. 
${\tilde{\nu}}_\mu$ / ${\tilde{\nu}}_\tau$ masses of a few hundreds GeV/c$^2$
are probed and excluded for relatively small couplings \cite{aleph2001}.
Much weaker limits can be extracted for ${\tilde{\nu}}_e$.

Additional heavy neutral bosons are predicted by many GUT models \cite{gut}.
LEP data at the $Z^0$ peak energy are sensitive to the
mixing angle $\theta_{ZZ'}$ of the $Z^0$ with a possible heavier $Z'$,
while LEP2 data are sensitive to its mass $m_{Z'}$.
Fits using all hadronic and leptonic cross sections and 
leptonic forward-backward asymmetries are consistent with no extra $Z'$. 
DELPHI results \cite{delphiz} are shown in 
Fig.~\ref{figlimits}~({\it right plot}).
The upper limits on the mixing angle $|\theta_{ZZ'}|$ are
about $2$~mrads. Assuming $\theta_{ZZ'} = 0$ the combined LEP data
have been fitted to determine $95 \%$ C.L. lower
limits on the $Z'$ mass. The resulting limits are $678 / 463 / 436$~GeV
respectively for $E(6)$ $\chi / \psi / \eta$ model, $800$~GeV for $L-R$ model
and $1890$~GeV for $SSM$ \cite{lepew}. 

Recently an idea has been proposed that Quantum Gravity scale could be as low as
$\approx 1$~TeV if gravitons propagate in large compactified extra dimensions,
while other particles are confined to the ordinary 3+1-dimensional world
\cite{aadd}. 
Gravity would be modified at distances of the order of the {\it size} of the
extra dimensions. This would solve the hierarchy problem, 
that is the striking difference between the electroweak scale 
($\approx 10^3$~GeV) and the Planck scale ($\approx 10^{19}$~GeV).
Existing gravity measurements stop at about $1$~mm, leaving room 
for new physics below this scale. New effects could be
within the reach of present and future colliders.
Virtual graviton exchange would modify
the fermion pair cross sections through interference terms proportional to
$\lambda / M_s^4$, where $\lambda$ is a parameter of ${\cal O}(1)$ depending on
the details of the theory and $M_s$ is a mass scale related to the Planck
scale in the $(4+n)$-dimensional space \cite{hewriz}. 
Pure graviton exchange would lead to terms of order $\lambda^2 / M_s^8$.
Bhabha scattering has the maximum sensitivity to low scale gravity effects, due
to interference with the dominant $t$-channel photon exchange. 
ALEPH \cite{aleph2001}, L3 \cite{l3_extra} and OPAL \cite{opal_extra} 
have analyzed all LEP2 Bhabha data and obtained lower limits on $M_s$ 
at about $1$~TeV. Such limits are derived by setting $\lambda = \pm 1$ to
account for positive or negative interference, with $M_s$ defined
according to \cite{hewriz}, and are shown in Table \ref{mslimits}.

In the near future each experiment is expected to finalize its data analyses
while the LEP working group should find a final agreement on 
exactly how to do 
the combinations (definitions, method, common uncertainties) and 
which results to combine.
In particular Bhabha measurements are still in the waiting-list. 
They are the most sensitive ones for many indirect searches, 
but in this case
theoretical uncertainties could be a serious limitation for the final results.

\begin{table}[htb]
\centering
\begin{tabular}{|c|c|c|} 
\hline
       & $\lambda = +1$ & $\lambda = -1$ \\
\hline
ALEPH  &  1.18          &  0.80          \\
\hline
L3     &  1.06          &  0.98          \\
\hline
OPAL   &  1.00          &  1.15          \\
\hline
\end{tabular}
\caption{\small 
Preliminary $M_s$ lower limits (95\% c.l.) in TeV from Bhabha analyses of
LEP experiments.}
\label{mslimits}
\end{table}
\noindent{\small \bf Acknowledgment}\\
I wish to thank all the organizers for the interesting conference
and the nice week we spent together.

\end{document}